\documentclass[sigconf]{acmart}

\usepackage[english]{babel}
\usepackage[utf8x]{inputenc}
\usepackage{dsfont}
\usepackage[TS1,T1]{fontenc}

\usepackage{booktabs} 
\usepackage{xcolor}
\usepackage{balance}

\usepackage{algorithm}
\usepackage{algorithmic}
\usepackage{amsmath}
\usepackage{listings}
\usepackage{mathtools}
\usepackage{graphicx}
\usepackage[position=bottom]{subfig}
\usepackage{float}
\usepackage{xspace}

\usepackage{color}

\newcommand{\eat}[1]{}
\newcommand*{\rf}{\textsf{Ranking Facts}\xspace}

\begin{document}
\title{A Nutritional Label for Rankings}

\author{Ke Yang}
\affiliation{
  \institution{Drexel University}
}
\email{ky323@drexel.edu}

\author{Julia Stoyanovich}\authornote{This work was supported in part by NSF Grant No. 1741047.}
\affiliation{
  \institution{Drexel University}
}
\email{stoyanovich@drexel.edu}

\author{Abolfazl Asudeh}
\affiliation{
  \institution{University of Michigan}
}
\email{asudeh@umich.edu}

\author{Bill Howe}\authornote{This work was supported in part by NSF Grant No. 1740996.}
\affiliation{
  \institution{University of Washington}
}
\email{billhowe@uw.edu}

\author{HV Jagadish}\authornote{This work was supported in part by NSF Grant No. 1741022.}
\affiliation{
  \institution{University of Michigan}
}
\email{jag@umich.edu}

\author{Gerome Miklau}\authornote{This work was supported in part by NSF Grant No. 1741254.}
\affiliation{
  \institution{University of Massachusetts Amherst}
}
\email{miklau@cs.umass.edu}

\renewcommand\shortauthors{Yang, Stoyanovich, et al.}

\begin{abstract}

Algorithmic decisions often result in scoring and ranking individuals to determine credit worthiness, qualifications for college admissions and employment, and compatibility as dating partners.  While automatic and seemingly objective, ranking algorithms can discriminate against individuals and protected groups, and exhibit low diversity. Furthermore, ranked results are often unstable --- small changes in the input data or in the ranking methodology may lead to drastic changes in the output, making the result uninformative and easy to manipulate.  Similar concerns apply in cases where items other than individuals are ranked, including colleges, academic departments, or products.

In this demonstration we present \rf, a Web-based application that generates a  {\em ``nutritional label''} for rankings. \rf is made up of a collection of visual widgets
that implement our latest research results on fairness, stability, and transparency for rankings, and that communicate details of the ranking methodology, or of the output, to the end user. We will showcase \rf on real datasets from different domains, including college rankings, criminal risk assessment, and financial services. 

\end{abstract}

\copyrightyear{2018}
\acmYear{2018}
\setcopyright{acmcopyright}
\acmConference[SIGMOD'18]{2018 International Conference on Management of Data}{June 10--15, 2018}{Houston, TX, USA}
\acmPrice{15.00}
\acmDOI{10.1145/3183713.3193568}
\acmISBN{978-1-4503-4703-7/18/06}

\fancyhead{}  

\maketitle

\section{Introduction}
\label{sec:intro}

Algorithmic decisions often result in scoring and ranking individuals --- to determine credit worthiness, desirability for college admissions and employment, and compatibility as dating partners.  While automatic and seemingly objective, rankers can
discriminate against individuals and protected groups~\cite{CitronP14}, and exhibit low diversity~\cite{DBLP:conf/edbt/StoyanovichAM11}. Furthermore, ranked results are often unstable --- small changes in the input or in  the ranking methodology may lead to drastic changes in the output, making the result uninformative and easy to manipulate~\cite{gladwell,label}.  Similar concerns apply in cases where items other than individuals are ranked, including colleges, academic departments, and products.

Algorithmic decisions are produced by complex processes with many hidden assumptions, and are increasingly used outside of the original context for which they were intended. 
In response, developers, regulators and the public need to quickly determine the ``fitness for use'' of a given model or dataset, and to assess the methodology that was used to produce it. This motivates development of interpretability and transparency tools.

With the exception of recent machine learning results that enable interpretability of particular classes of algorithms ~\cite{DBLP:conf/kdd/Ribeiro0G16,DBLP:conf/kdd/Rudin14,DBLP:conf/icdm/WangRDLKM16}, recent scholarship on algorithmic accountability has primarily focused on enabling an analyst to retroactively verify particular properties rather than proactively exposing a standard suite of information.  Because algorithmic processes can be complex or secret, these methods rely on retrospective checks, using techniques like zero knowledge proofs~\cite{Kroll2017}, audits~\cite{Sandvig}, and reverse engineering~\cite{PerelElkinKoren}. These are valid methods of interrogation, but they put a significant burden on users.  The burden should instead be borne by the vendor who produced the result, who is in a better position to explain it.

In this work we develop an interpretability tool, \rf, that is based on the concept of a {\em nutritional label}.  We draw an analogy to the food industry, where simple, standardized labels convey information to consumers about the ingredients and production processes.  Short of setting up a chemistry lab, the consumer would otherwise have no access to this information.  Similarly, \rf explains ranked outputs to a user, with appropriately summarized information regarding the ranking process.

\begin{figure*}[t!]	
\centering
\includegraphics[width=\linewidth]{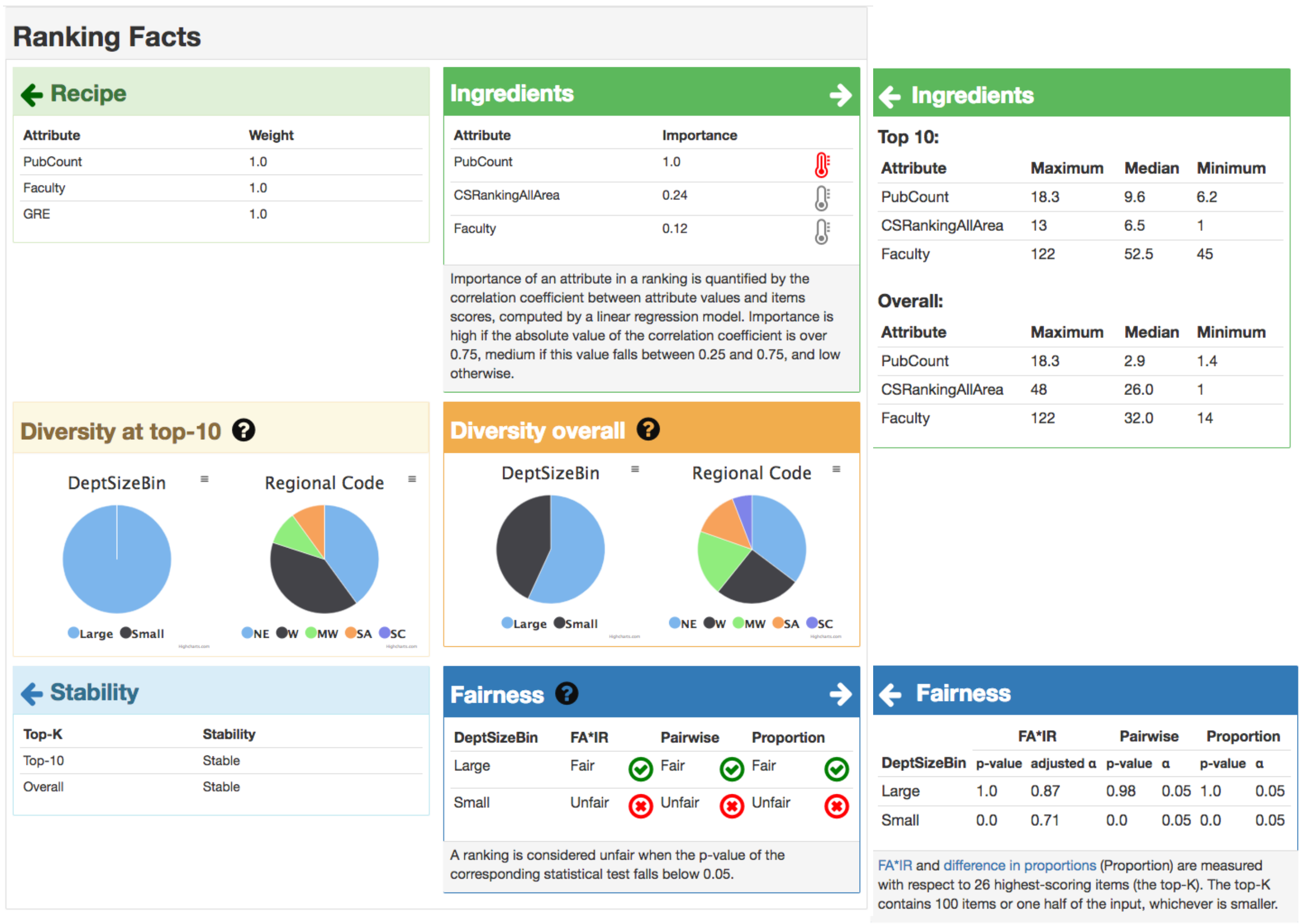}
     \caption{\rf for the CS departments dataset. The Ingredients widget (green) has been expanded to show the details of the attributes that strongly influence the ranking.  The Fairness widget (blue) has been expanded to show the computation that produced the fair/unfair labels.}
     \label{fig:label}
\end{figure*}

An example of the output produced by our tool is presented in Figure~\ref{fig:label}. It explains a ranked set of Computer Science departments.  The data was obtained from CS Rankings (\url{https://github.com/emeryberger/CSRankings}), augmented with attributes from the NRC (\url{http://www.nap.edu/rdp/}) dataset, see details in Section~\ref{sec:demo}. 

\rf is made up of a collection of visual widgets.  Each widget addresses an essential aspect of transparency and interpretability, and is based on our recent technical work on fairness and diversity~\cite{diversity,DBLP:conf/edbt/StoyanovichAM11,DBLP:conf/edbt/StoyanovichYJ18,DBLP:conf/ssdbm/YangS17}, transparency~\cite{label}, and stability (ongoing) in algorithmic rankers. 
We describe next how we explain rankings using the widgets (Section~\ref{sec:sys}) and then discuss demonstration scenarios (Section~\ref{sec:demo}) before concluding in Section~\ref{sec:conc}.

\section{Explaining rankings}
\label{sec:sys}

Figure~\ref{fig:label} presents \rf for CS department rankings.  The nutritional label consists of six widgets, each with an overview and a detailed view, which we now describe.  

\subsection{Recipe and Ingredients}

These two widgets help to explain the ranking methodology.  The {\sf Recipe} widget succinctly describes the ranking algorithm.  For example, for a linear scoring formula, each attribute would be listed together with its weight. The {\sf Ingredients} widget lists attributes most material to the ranked outcome, in order of importance.  For example, for a linear model, this list could present the attributes with the highest learned weights.  Put another way, the explicit intentions of the designer of the scoring function about which attributes matter, and to what extent, are stated in the {\sf Recipe}, while {\sf Ingredients} may show additional attributes associated with high rank.  Such associations can be derived with linear models or with other methods, such as rank-aware similarity in our prior work~\cite{DBLP:conf/edbt/StoyanovichAM11}.

The detailed {\sf Recipe} and {\sf Ingredients} widgets list statistics of the attributes in the {\sf Recipe} and in the {\sf Ingredients}: minimum, maximum and median values at the top-$10$ and over-all.

\subsection{Stability}
\label{sec:sys:stab}

The {\sf Stability} widget explains whether the ranking methodology is robust on this particular dataset.  An unstable ranking is one where slight changes to the data (e.g., due to uncertainty and noise), or to the methodology (e.g., by slightly adjusting the weights in a score-based ranker) could lead to a significant change in the output. This widget reports a stability score, as a single number that indicates the extent of the change required for the ranking to change.

\begin{figure}[t!]	
\centering
\includegraphics[width=3in]{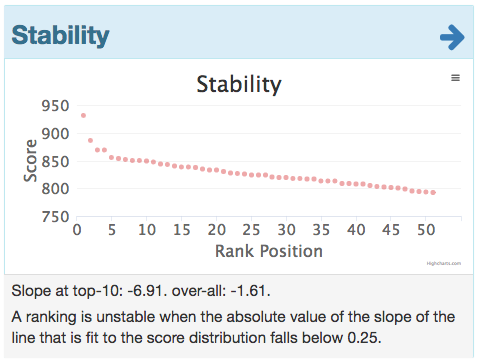}
     \caption{{\sf Stability}: detailed widget.}
     \label{fig:stability}
\end{figure}

As with the widgets above, there is a detailed {\sf Stability} widget to complement the overview widget.  An example is shown in Figure~\ref{fig:stability}, where the stability of the ranking is quantified as the slope of the line that is fit to the score distribution, at the top-$10$ and over-all.  A score distribution is unstable if scores of items in adjacent ranks are close to each other, and so a very small change in scores will lead to a change in the ranking.  In this example the score distribution is  considered unstable if the slope is 0.25 or lower.  Alternatively, stability can be computed with respect to each scoring attribute, or it can be assessed using a model of uncertainty in the data.

\subsection{Fairness}
\label{sec:sys:fair}

The {\sf Fairness} widget quantifies whether the ranked output exhibits statistical parity (one interpretation of fairness) with respect to one or more sensitive attributes, such as gender or race of individuals~\cite{DBLP:conf/ssdbm/YangS17}.  We denote one or several values of the sensitive attribute as a protected feature.  For example, for the sensitive attribute {\sf gender}, the assignment {\sf gender=F} is a protected feature.

A variety of fairness measures have been proposed in the literature~\cite{DBLP:journals/datamine/Zliobaite17}, but none are directly applicable to rankings. One typical measure compares the proportion of members of a protected group (e.g., female gender or minority race) who receive a positive outcome to their proportion in the overall population.  For example, if the dataset contains an equal number of men and women, then among the individuals invited for a job interview, one half should be women.  A measure of this kind can be adapted to rankings by quantifying the proportion of members of a protected group in some selected set of size $k$ (treating the top-$k$ as a set).

In~\cite{DBLP:conf/ssdbm/YangS17}, we proposed a generative method to describe rankings that meet a particular fairness criterion (fairness probability $f$) and are drawn from a dataset with a given proportion of members of a binary protected group ($p$). This method was used in FA*IR~\cite{DBLP:conf/cikm/ZehlikeB0HMB17} to quantify fairness in every prefix of a top-$k$ list.  In our follow-up work (working paper), we are developing a pairwise measure that directly models the probability that a member of a protected group is preferred to a member of the non-protected group.

Let us now return to the {\sf Fairness} widget in Figure~\ref{fig:label}.  We select a binary version of the department size attribute {\sf DeptSizeBin} from the CS departments dataset as the sensitive attribute, and treat both values (``large'' and ``small'') as protected features.  The summary view of the {\sf Fairness} widget in our example presents the output of three fairness measures: FA*IR~\cite{DBLP:conf/cikm/ZehlikeB0HMB17}, proportion~\cite{DBLP:journals/datamine/Zliobaite17}, and our own pairwise measure.  All these measures are statistical tests, and whether a result is fair is determined by the computed p-value.  The detailed {\sf Fairness} widget provides additional information about the tests and explains the process.

\subsection{Diversity}
\label{sec:sys:div}

Fairness is related to diversity: ensuring that different kinds of objects are represented in the output of an algorithmic process~\cite{diversity}.  Diversity has been considered in search and recommender systems, but in a narrow context, and was rarely applied to profiles of individuals.  We are currently working on defining diversity measures for ranked outputs, based on our work in~\cite{diversity,DBLP:conf/edbt/StoyanovichAM11}. 

The {\sf Diversity} widget  shows diversity with respect to a set of demographic categories of individuals, or a set of categorical attributes of other kinds of items~\cite{diversity}.  The widget displays the proportion of each category in the top-$10$ ranked list and over-all, and, like other widgets, is updated as the user selects different ranking methods or sets different weights.  In our example in Figure~\ref{fig:label}, we quantify diversity with respect to department size and to the regional code of the university.  By comparing the pie charts for top-$10$ and over-all, we observe that only large departments are present in the top-$10$.

\section{Demonstration Scenarios}
\label{sec:demo}

\begin{figure*}[t!]	
\centering
\includegraphics[width=5.5in]{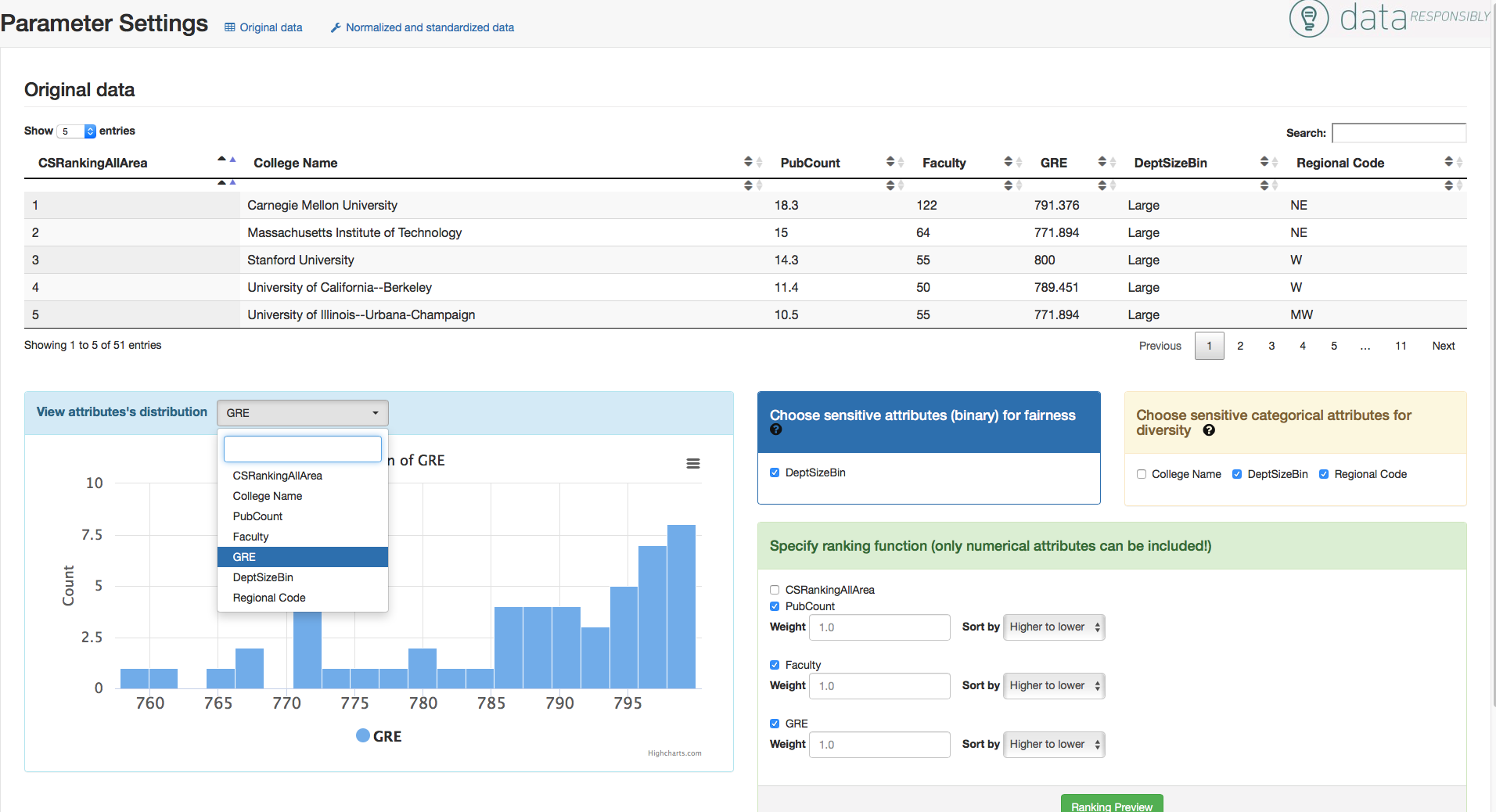}
     \caption{Scoring function design.}
     \label{fig:design}
\end{figure*}

We will demonstrate the utility of \rf using three real-world data sets, considering several ranking functions for each. 

(1) CS departments: CS Rankings (CSR) (\url{https://github.com/emeryberger/CSRankings}), with additional attributes from the NRC assessment dataset (\url{http://www.nap.edu/rdp/}).  This dataset has the following attributes: {\sf PubCount} (CSR) computes the geometric mean of the adjusted number of publications in each area by institution, {\sf Faculty} (CSR) is the number of faculty in the department, {\sf GRE} (NRC) is the average GRE scores (2004-2006), and {\sf Region} (NRC) is one of NE, MW, SA, SC, W regions in the US.  We have been using this dataset in our examples throughout this paper.  

(2) Criminal risk assessment: a dataset collected and published by ProPublica as part of their investigation into racial bias in criminal risk assessment software called COMPAS (\url{https://github.com/propublica/compas-analysis}). The dataset contains demographics, recidivism scores produced by COMPAS, and criminal offense information for 6,889 individuals. 

(3) Credit and loans: the German Credit dataset from the UCI Machine Learning Repository (\url{https://archive.ics.uci.edu/ml/datasets/}), with demographic and financial information on 1000 individuals.

The demo user has the option to choose one of these datasets, or to upload one of their own (as a fully populated table in CSV format).  Next, they can choose a ranking method from pre-populated options, or they can rank using scoring function of their own specification. If they make the latter choice, the system provides assistance.  Figure~\ref{fig:design} presents a portion of the scoring function design view.  Here, the user can decide whether to work with raw data or to normalize and standardize the attributes (checkbox at the top-left of Figure~\ref{fig:design}).  The system generates a preview of the data, and allows the user to plot the distribution of values of each attribute as a histogram (shown here for the attribute {\sf GRE}).  

The bottom-right portion of Figure~\ref{fig:design} contains the attribute selection areas: at least one categorical attribute must be chosen as the sensitive attribute.  \rf will evaluate fairness with respect to every value in the domain of this attribute, and is currently limited to binary attributes.  Finally, the user selects at least one numerical attribute for the scoring function, and assigns a weight to this attribute.  Scoring attribute selection and weight assignment is based on a user's a priori judgment regarding item quality, but can be informed by the range and distribution of values for a given attribute.  When scoring attributes are selected, the user will preview the ranking, and will then either refine it, or go on to generate \rf such as that in Figure~\ref{fig:label}.

The demo presenter will inspect the resulting nutritional label together with the user, guiding the user in exploring selected widgets in detail.  For example, we may notice that many attributes in the {\sf Recipe} do not coincide with those that most impact the ranked outcome in the {\sf Ingredients}, as is the case in Figure~\ref{fig:label}: attribute {\sf GRE} is one of the scoring attributes, but it does not correlate with the ranked outcome.  Inspecting the detailed {\sf Recipe} widget, we observe that the range of values and the median for {\sf GRE} are very similar in the top-$10$ and overall, supporting the finding that {\sf GRE} does not play an important part in the ranking.

\section{Take-Away Messages}
\label{sec:conc}

In this demonstration we presented \rf, a system for producing nutritional labels that explain rankings.  To the best of our knowledge, our tools are the first to consider interpretability for ranked outputs.  \rf is implemented in Python, and is modular and easy to extend.  Our tool is available at \url{http://demo.dataresponsibly.com/rankingfacts/}.

The tool is based on the latest research by the authors and others, and reflects known limitations of the state of the art.  We are actively working on defining group fairness measures that go beyond binary categories (e.g., can be applied to ethnicity, not only to gender), and will incorporate these into the tool when available. We are also working on extending \rf to support richer scoring function design functionality.  For example, we plan to include methods that help the user mitigate lack of fairness and diversity by suggesting modified scoring functions.

\balance

\bibliographystyle{ACM-Reference-Format}
\bibliography{demorefs}

\end{document}